\documentclass[twocolumn, showpacs,preprintnumbers,%
amsmath,amssymb,prc]{revtex4}
\begin{document}
\preprint{APS/123-QED}
\title{$\omega$ Production in $p\,p$ Collisions}
\author{G. Ramachandran}
 \affiliation{Indian Institute of Astrophysics,
Koramangala, 
Bangalore, 
560034}
\author{M. S. Vidya}
 \affiliation{School of Physical Sciences, Jawaharlal Nehru University, New
Delhi, 110067}
\author{P. N. Deepak}
 \affiliation{Institute f\"{u}r Kernphysik,
Forschungszentrum 
J\"ulich,
D-52425 J\"{u}lich, Germany}
\author{J. Balasubramanyam}
 \affiliation{K. S. Institute of Technology,
Bangalore, 560062}
\author{Venkataraya}
 \affiliation{Vijaya College, Bangalore, 560011}
\date{\today}

\begin{abstract}
A model-independent irreducible tensor formalism which has been
developed earlier to analyze measurements of $\vec{p}\vec{p}\to
pp\,\pi^\circ$, is extended to present a theoretical discussion of
$\vec{p}\vec{p}\to p p \,\omega$ and of $\omega$  polarization in $pp\to pp\,
\vec{\omega}$ and in $p\vec{p}\to p p\,\vec{\omega}$. The recent
measurement of unpolarized differential cross section for $p p \to p p \,\omega$ 
is analyzed using this theoretical formalism.
\end{abstract}
\pacs{13.75.Cs, 13.88.+e, 21.30.-x, 24.70.+s,
25.40.Ve}
\maketitle

Experimental study of meson production in $NN$ collisions has
attracted considerable interest during the last decade
and a half. The early measurements of total cross-section
\cite{1} for pion production were found surprisingly to be more
than a factor of 5 than the theoretical predictions \cite{2}.
At c.m. energies close to threshold, the relative kinetic
energies between the particles in the final state are small and
an analysis involves, therefore, only a few partial
waves. On the other hand, a large momentum transfer is involved
when an additional particle is produced in the final state,
thus making the reaction sensitive  to the
features of the $NN$ interaction at short distances
where the nucleons start to overlap. When a heavier meson
like $\omega$ is produced, the overlapping region
corresponds  \cite{3} to a distance of about $0.2fm$.
It is also known that the short range part of the $NN$
interaction is dominated by the $\omega$ exchange \cite{4}.
 Consequently, a variety of theoretical models have been proposed \cite{5} not
only to bridge the gap between theory and experiment, but also to test
results of QCD based  discussions of the $NN$ interaction.
According to the OZI rule \cite{6}, $\phi$ 
production relative to  $\omega$ production is suppressed in the
absence of strange quarks in the initial state. This 
ratio $R$ has been measured \cite{7}, in view of the
dramatic  violations \cite{8} observed in $\bar{p}p$ collisions,
and compared to the theoretical estimate 
\cite{9} of $4.2 \times 10^{-3}$
after correcting for the available phase space. 
We may  refer  \cite{10} for modifications of the
rule. Apart from looking for the strange quark content
of the nucleon in the initial state, attention has also been
focused on resonance contributions 
\cite{11,12,13} to vector meson production in $NN$
collisions.  The constituent quark models \cite{14}
predict highly excited $N^*$ states which  have not been 
seen in $\pi N$ scattering. This `` missing resonance problem "
\cite{15} has also catalyzed the experimental study of $\omega$
meson production in the hope that the missing resonances may couple more
strongly or even exclusively to the $\omega N$ channel in comparison to
the $\pi N$ channel, although $ \omega N $ decay modes of resonances 
have not been observed \cite{16}. Also the cross-sections of vector meson production
enter as inputs into transport  models for dilepton emission  in heavy 
ion collisions which may in turn be used to study the
off-shell $\omega$ production and medium modifications of the widths and
masses of the resonances \cite{13}.

Meson production in $NN$ collisions involves also spin
state transitions of the $NN$ system, which do not
occur in elastic $NN$ scattering. In $pp\to pp \pi^0 $,
for example, the transition of the $pp$ system at
threshold is from an initial spin triplet to a
final spin singlet state $(^3P_0 \to \ ^{1}S_0)$. 
 Rapid advances in experimental technology have led today to  high
precision measurements of spin observables \cite {17} at several
energies up to 400 MeV, employing beams of polarized protons on
polarized proton targets.  Conclusive theoretical
interpretation of all these data have remained
elusive, although  the model calculations appear to do better
in the case of charged pion production as compared
to the neutral pion production and the agreement even
there seemed to deteriorate increasingly at higher energies.
It has been pointed out both by Moskal {\it{et al.}}
\cite{5} and Hanhart \cite{5}, that the extensive experimental
information available comes with a drawback that
``apart from rare cases, it is difficult to extract a
particular piece of information from the data''.

A model-independent irreducible tensor formalism
\cite{18} which has been developed to analyze
measurements on $\vec{p}\vec{p}\to pp\pi^0$  at the complete
kinematical double differential  level, was recently \cite{19}
made use of to estimate empirically the initial
singlet and triplet state contributions to the differential
cross-section using the experimental results of Meyer
\textit{et al.} \cite{17}.  The above theoretical
formalism leads, on integration, to the relation derived earlier
by Bilenky and Ryndin \cite{20} for the total
cross-sections. It was also shown \cite{21} how the irreducible tensor
formalism could be utilized to effect spin filtering,
in general, for any scattering or reaction process employing
polarized beams of particles with arbitrary spin $s_b$  on polarized
targets with arbitrary spin $s_t$.
The production of a heavy meson like $\omega$, at and
near threshold in $\vec{p}\vec{p}$ collisions  allows
us to study additional spin dependent features of $NN$
interactions at much shorter distances. Unlike the pion which is
spinless, the $\omega$  has spin 1 which permits us to make
observations with regard to its spin state also apart from
measuring the angular distributions in polarized beam and polarized
target experiments.  Experimental data on total \cite{22} and
differential \cite{23} cross-sections for $pp \to pp \omega$ have
 already  been published and proposals are underway
\cite{5,24} to study heavy meson production in $NN$
collisions using polarized  beams and targets at COSY.

The purpose of the present paper is to extend the earlier work \cite{18,19}
 on the model independent approach based on
irreducible tensor techniques, to study the spin state of the meson in
$pp\to pp\vec{\omega}$ and $p\vec{p}\to pp\vec{\omega}$
as well as the double differential cross section in
the proposed  polarized beam and polarized target experiments.

Let $ \boldsymbol{p}_i$ denote the initial c.m. momentum,  $\boldsymbol{q}$ 
the momentum of the meson produced with spin parity $s^{\pi} $ and
${\boldsymbol{p}}_f$ the relative momentum,  $ (1/2)(\boldsymbol{p}_1 -\boldsymbol{p}_2)$
between the two nucleons with c.m. momenta
$\boldsymbol{p}_1$ and $ \boldsymbol{p}_2 $ in the final state.
The double differential cross section for meson production in c.m. may be written as
\begin{equation}
\label{ddcs1}
\text{d}^2 \sigma = \frac{2\pi D}{v}
\text{Tr}(\boldsymbol{T}\rho^i \boldsymbol{T}^\dagger),
\end{equation} where   $D$  denotes  the final three particle density
of states, $ \boldsymbol{T} $ denotes the  on-energy-shell
transition matrix and $ \boldsymbol{T}^{\dagger}$  its
hermitian conjugate, $v = 4 |{\boldsymbol{p}}_i|/E $ at c.m
energy $E$ and $\rho^i$ denotes the initial   spin density matrix,
\begin{equation}
\rho^i = {\textstyle{\frac{1}{4}}}(1+\boldsymbol{\sigma}_1
\cdot \boldsymbol{P})(1+\boldsymbol{\sigma}_2 \cdot \boldsymbol{Q}),
\end{equation}
if $\boldsymbol{P}$ and $\boldsymbol{Q}$  denote
respectively the beam and target polarizations. Notation
$\sigma(\xi,\boldsymbol{P},\boldsymbol{Q})$
is used in \cite{17} to denote \eqref{ddcs1}.
If $s_i$  and $s_f$  denote the initial
and final spin states of the $NN$ system, the initial and final
channel spins for the reaction are $s_i$ and $ S $
respectively, where $S$ can
assume values $ S=|s_f - s|, \ldots (s_f+s) $.  Making
use of the irreducible tensor operator techniques
introduced in \cite{25}, we may express $\boldsymbol{T}$ in the
operator form
\begin{eqnarray}
\label{tm}
\boldsymbol{T} & = &
\sum_{\alpha}\sum_{\lambda=|s_f-s_i|}^{(s_f+s_i)}
\sum_{\Lambda = |S-s_i|}^{(S+s_i)} \nonumber \\ & & \times(( S^s(s,0) \otimes
S^{\lambda}(s_f,s_i))^{\Lambda} \cdot \mathcal{T}^{\Lambda}(\alpha,\lambda)),
\end{eqnarray}
where $\alpha = ( S,s_f,s_i) $ denotes collectively the
spin variables.  The irreducible tensor amplitudes
$ \mathcal{T}_{\nu}^{\Lambda}(\alpha,\lambda) $ of rank $\Lambda$, which
characterize the reaction, are given by
\begin{eqnarray}
\label{TLnu}
\mathcal{T}^{\Lambda}_\nu(\alpha,\lambda)&=&W(s s_f \Lambda s_i ;S
\lambda)[\lambda]\sum_{\beta} \sum_{j} T_{\alpha,\beta}^{j}
  \,W(s_il_iSL;j \Lambda) \nonumber \\
& & \times (( Y_l(\boldsymbol{\hat{q}})\otimes
Y_{l_f}(\boldsymbol{\hat{p}}_f))^L   \otimes
Y_{l_i}(\boldsymbol{\hat{p}}_i))_{\nu}^{\Lambda},
\end{eqnarray}
in terms of the partial wave amplitudes 
\begin{eqnarray}
T_{\alpha,\beta}^{j} & = &(4 \pi)^3 (-1)^{L+l_i+s_i-j} 
[j]^2[S][s]^{-1}[s_f]^{-1} \nonumber \\  & & \times
\langle((ll_f)L(ss_f)S)j||T||(l_is_i)j \rangle,
\end{eqnarray}
which depend on  $E$ and invariant mass  $W$ of the final 
$NN$ system.  Total angular momentum $j$ is 
conserved and $\beta = (l,l_f, L, l_i) $ 
denotes collectively the orbital angular momentum $l$
of the emitted  meson, the initial and final relative orbital angular
momenta $l_i$  and $l_f$ of the $NN$ system and the
total orbital angular momentum $L$
in the final state, which takes values $L = |l_f-l|, \ldots ,(l_f+l)$. It may 
be noted that our coupling of angular momenta in the final state differs from 
that used by  Meyer \emph{et al.,} \cite{17} in the case of the production 
of a meson with spin $s=0$. 
The notation $[\lambda]=\sqrt{2 \lambda+1}$ is used
apart from  standard notations \cite{26}. The above formalism is
readily extendable to  arbitrary charge states of hadrons in $NN \to NNx $ 
where $x$ represents a meson with isospin $I_s$,   if we identify
\begin{eqnarray}
T_{\alpha,\beta}^{j}&=& \sum_{I_i,I_f}
C({\textstyle{\frac{1}{2}}}  {\textstyle{\frac{1}{2}}}
I_i;\nu_1^i \nu_2^i \nu_i)C({\textstyle{\frac{1}{2}}}  {\textstyle{\frac{1}{2}}}
I_f;\nu_1^f \nu_2^f \nu_f)\nonumber \\ && \times
C(I_f I_s I_i;\nu_f \nu_s \nu_i)T_{\alpha,\beta}^{I_fI_i \,j},
\label{iso}
\end{eqnarray}
where $I_i$  and $I_f$  denote respectively the initial and final
isospin quantum numbers of the $NN$ system. 
We have $I_i=I_f=\nu_i=\nu_f=1$, here with $I_s=0$.   Pauli exclusion
principle and parity conservation restrict the summations 
in \eqref{tm} and \eqref{TLnu} to terms satisfying
$(-1)^{l_i + s_i + I_i} =  -1 = (-1)^{l_f + s_f + I_f} ;
(-1)^{l_i}  =\pi\,(-1)^{l_f+l}$.
Thus, the contributing partial waves in $ pp\to pp \omega $
at and  near threshold may be taken as shown in Table~\ref{amplitudes}, 
 where we use the same notations as in \cite{17} viz, $S,P,D \ldots $ for $l_i, l_f 
= 0,1, 2,\ldots $ and $s,p,d \ldots $ for $l = 0, 1, 2,\ldots $. We use 
$\cal S, \cal P, \cal D,\ldots $ for $L = 0, 1, 2,\ldots $  in the final state. 
We now express
\begin{table}[t]
\caption{\label{amplitudes}The irreducible tensor amplitudes and the
partial wave contributions to $pp \to pp \omega$ close to threshold }
\begin{ruledtabular}
\begin{tabular}{lccccccccccc}
${\cal  T}_{\nu}^{\Lambda}(\alpha,\lambda)$ &
$l_f$ & $l$ & $L$ & $s_f$ & $S$ & $j$ & $l_i$ & $s_i $ &
$ T_{\alpha,\beta}^{j}$ & Initial & Final \\
&&&&&&&&&&$pp$ state &$pp\omega$ state \\ \hline 
${\cal  T}_{\nu}^{1}(101;1)$ &0&0&0&0&1&1&1&1& $T_{101;0001}^{1}$ &
$^3P_1$ & $(^1Ss)^3\mathcal{S}_1$\\  $
{\cal  T}_{\nu}^{1}(100;0)$&0&1&1&0&1&0&0&0&$T_{100;1010}^{0}$&
$^1S_0$& $(^1Sp)^3\mathcal{P}_0$\\
&0&1&1&0&1&2&2&0&$T_{100;1012}^{2}$&
$^1D_2$ & $(^1Sp)^3\mathcal{P}_2$\\  
${\cal  T}_{\nu}^{1}(110;1)$&1&0&1&1&1&0&0&0&$T_{110;0110}^{0}$&$^1S_0
$&$(^3Ps)^3\mathcal{P}_0$\\
&1&0&1&1&1&2&2&0&$T_{110;0112}^{2}$&$^1D_2$&$(^3Ps)^3\mathcal{P}_2$\\  
${\cal  T}_{\nu}^{2}(210;1)$
&1&0&1&1&2&2&2&0&$T_{210;0112}^{2}$&$^1D_2$
&$(^3Ps)^5\mathcal{P}_2$ \\ 
\end{tabular}
\end{ruledtabular}
\end{table}
\begin{equation}
\rho^i = \sum_{s_i,s_i'=0}^{1} \sum_{k=|s_i -s_i'|}^{(s_i + s_i')}
(S^k(s_i,s_i') \cdot I^k(s_i,s_i')),
\end{equation}
in terms of irreducible tensor operators $S_{\nu}^k(s_i,s_i')$
and the initial polarization tensors
\begin{equation}
\label{Iknu}
  I_{\nu}^k(s_i,s_i') = \sum_{k_1,k_2=0}^{1} 
F \; ( P^{k_1} \otimes Q^{k_2})_{\nu}^{k},
\end{equation}
of rank k, using the notations $ P_{0}^{0}=Q_{0}^{0}=1$
and $P_{\nu}^{1}, Q_{\nu}^{1}$   to denote the spherical components of
$\boldsymbol{P},\boldsymbol{Q}$  respectively and the factor
\begin{equation}
\label{wig}
F ={\textstyle{\frac{1}{2}}}
(-1)^{k_1+k_2-k}[k_1][k_2][s_i'] \left\{ \begin{matrix}
{\textstyle{\frac{1}{2}}} & {\textstyle{\frac{1}{2}}} & s_i \\[.2cm]
{\textstyle{\frac{1}{2}}} & {\textstyle{\frac{1}{2}}}
& s_i'\\[.2cm] k_1 &   k_2  & k
\end{matrix} \right \}.
\end{equation}

Using known properties \cite{25} of the irreducible tensor operators
and standard Racah techniques, we have 
\begin{equation}
\label{full-dcs}
\text{d}^2 \sigma = \sum_{\alpha,\alpha',\Delta,k} G\;\; 
(I^k(s_i,s_i') \cdot \mathcal{B}^k(s_i,s_i')),
\end{equation}
in terms of the bilinear irreducible tensors
\begin{align}
\mathcal{B}_{\nu}^{k}&(s_i,s_i')=  \frac{2 \pi D}{v} ({\cal  T}^{\Lambda}
(\alpha,\lambda) \otimes {\cal  T}^{\dagger  \Lambda'}(\alpha',\lambda'))_{\nu}^k,
\end{align}
of rank $k$ and the geometrical factors
\begin{align}
G= \delta_{s_f s_f'} [s_f]^2[s_i][s]^2
(-1)^{\lambda+\lambda'+\Lambda'}[\lambda][\Lambda][\lambda']
[\Lambda'] \nonumber \\
\times W(s \lambda \Lambda' k;\Lambda \lambda') 
W(s_ i'ks_f \lambda ;s_i \lambda'),
\end{align}
where ${\cal T}_{\nu}^{\dagger \Lambda}(\alpha,\lambda)$
and the complex conjugates ${\cal  T}_{\nu}^{\Lambda}
(\alpha,\lambda)^*$ of \eqref{TLnu} are related
through ${\cal  T}_{\nu}^{\dagger \Lambda}
(\alpha,\lambda) =(-1)^{\nu}{\cal  T}_
{-\nu}^{\Lambda}(\alpha,\lambda)^* $ and  $\Delta =
(\lambda,\lambda',\Lambda,\Lambda')$.

Defining the partial contributions to $
\text{d}^2 \sigma$ through $ \text{d}^2 \sigma = \sum_{s_i,s_i'}
\text{d}^2 \sigma (s_i,s_i')$ and using \eqref{Iknu}, we have
\begin{eqnarray}
\label{d00}
\text{d}^2 \sigma(0,0)&=& \text{d}^2\sigma_{0}{\textstyle\frac{1}{4}}(1-
\boldsymbol{P} \cdot \boldsymbol{Q})[1+\sqrt{3} {A}_{0}^{0}(11)],\\
\label{d11}
\text{d}^2 \sigma(1,1)&=& \text{d}^2
\sigma_{0}[{\textstyle\frac{1}{4}}(3+\boldsymbol{P}\cdot \boldsymbol{Q})
(1-{\textstyle{\frac{1}{\sqrt{3}}}}A_{0}^{0}(11))
  \nonumber \\ & &+{\textstyle\frac{1}{2}}(( 
\boldsymbol{P}+\boldsymbol{Q})\cdot ( {\boldsymbol{A}}(10)+{\boldsymbol{A}}(01)))
\nonumber\\ & &+ (( P^1 \otimes Q^1)^2 \cdot A^2(11))] ,\\
\label{d22}
\text{d}^2 \sigma(1,0) & + & \text{d}^2\ \sigma(0,1) = \text{d}^2
\sigma_{0} \nonumber \\ & & \times[{\textstyle\frac{1}{2}}
(( \boldsymbol{P}-\boldsymbol{Q})\cdot ( {\boldsymbol{A}}(10)-{\boldsymbol{A}}(01)))
\nonumber \\ & & +((P^1 \otimes Q^1)^1 \cdot A^1(11))],
\end{eqnarray}
which add up to give \eqref{full-dcs} in the form
\begin{eqnarray}
\label{dcs2}
\text{d}^2 \sigma &=& \text{d}^2 \sigma_0[1+ \boldsymbol{P} \cdot
\boldsymbol{A}(10) +\boldsymbol{Q} \cdot \boldsymbol{A}(01) \nonumber\\
& &+\sum_{k=0}^{2} ((P^{1}\otimes Q^{1})^k \cdot A^k(11))],
\end{eqnarray}
where the unpolarized double differential cross
section
\begin{equation}
\label{dcs0}
\text{d}^2 \sigma_0= \frac{1}{4}\sum_{\alpha,\lambda,\Lambda}
(-1)^{\Lambda}[s_f]^2[s]^2[\Lambda] {\cal B}^{0}_{0}(s_i,s_i),
\end{equation}
is denoted as $\sigma_0(\xi)$ in \cite{17}. The beam, target analyzing powers 
${\boldsymbol{A}}(01),{\boldsymbol{A}}(10)$ are represented by the irreducible 
tensors $A^1_{\nu}(10)$, $A^1_{\nu}(01)$ respectively and the spin correlations 
by $A^{k}_{\nu}(11)$ of rank $k=0,1,2$. We have
\begin{equation}
\label{ak12}
\text{d}^2 \sigma_0 A_{\nu}^{k}(k_1k_2)= \sum_{\alpha,\alpha',\Delta} F \;G\;\;
\mathcal{B}_{\nu}^{k}(s_i,s_i').
\end{equation}

Our $A_{\nu}^{k}(k_1k_2)$ are given, in terms of the notations of  Meyer \emph{et al.,}
\cite{17}, by
\begin{eqnarray}
A^{1}_{0}(10) &=& A_{z0}(\xi), \  A^{1}_{0}(01) =A_{0z}(\xi),\nonumber\\
A^{1}_{\pm 1}(10)&=&\mp {\textstyle\frac{1}{\sqrt{2}}}[A_{x0}(\xi)\pm 
iA_{y0}(\xi)],\nonumber\\
A^{1}_{\pm 1}(01) &=& \mp {\textstyle\frac{1}{\sqrt{2}}}[A_{0x}(\xi)\pm 
iA_{0y}(\xi)],\nonumber\\
A^0_{0}(11)&=&-{\textstyle\frac{1}{\sqrt{3}}}[A_{\Sigma}
(\xi)+A_{zz}(\xi)],\nonumber \\
A^1_{0}(11)&=&-{\textstyle \frac{i}{\sqrt{2}}}A_{\Xi}(\xi),\\
A^1_{\pm 1}(11)&=&{\textstyle \frac{1}{2}}[(A_{xz}
(\xi)-A_{zx}(\xi))\pm i(A_{yz}(\xi)-A_{zy}
(\xi))],\nonumber\\
A^2_{0}(11)&=&{\textstyle\frac{1}
{\sqrt{6}}}[2 A_{zz}(\xi)-A_{\Sigma}(\xi)],\nonumber\\
A^2_{\pm 1}(11)&=& \mp {\textstyle \frac{1}{2}}
[(A_{xz}(\xi)+A_{zx}(\xi))\pm i(A_{yz}(\xi)+A_{zy}(\xi))],\nonumber\\
A^2_{\pm 2}(11)&=& {\textstyle \frac{1}{2}}[A_{\Delta}
(\xi)\pm i(A_{xy}(\xi)+ A_{yx}(\xi))],\nonumber
\end{eqnarray}
where $A_{ij}(\xi),\, i,j=0,x,y,z$ are the same as in Eq.(4) of \cite{17} 
and $A_{\Sigma}, A_{\Delta}, A_{\Xi}$ are defined by Eq.(5) of \cite{17}.

At a $ \vec{p} \vec{p}$ facility similar to  PINTEX at IUCF, but 
with sufficiently high energies $E$,  it should,
therefore, be possible to determine \eqref{d00} to \eqref{d22} 
individually apart from \eqref{dcs2} and \eqref{dcs0}. 
It is interesting to note from
Table~\ref{amplitudes} that only the
$\mathcal{T}^{1}_{\nu}(101;1)$
from the initial state $ ^3P_1$ contribute to 
\eqref{d11}  and hence
$ |T^1_{101;0001}|^2$ can be determined empirically, while \eqref{d22} gets 
contributions to the interference of $T^1_{101;0001}$ with
all the other five singlet amplitudes, which by themselves
determine  \eqref{d00}. Moreover we  note that 
$\text{d}^2 \sigma_0$ given by \eqref{dcs0} may itself be decomposed into 
$\sum_{s_i,m_i}\;^{2s_i+1}(\text{d}^2 \sigma_0)_{m_i} $, where
\begin{align}
\label{d200}
^1(\text{d}^2 \sigma_0)_0&= {\textstyle\frac{\text{d}^2 \sigma_{0}}{4}} 
[1+\sqrt{3} {A}_{0}^{0}(11)],\\ 
\label{d210}
^3(\text{d}^2 \sigma_0)_0 &= {\textstyle\frac{\text{d}^2 \sigma_{0}}{4}}
[1-{\textstyle{\frac{1}{\sqrt{3}}}}A_{0}^{0}(11)-
{\textstyle\frac{2 \sqrt{2}}{\sqrt{3}}}A_{0}^{2}(11)],\\
\label{d211}
^3(\text{d}^2 \sigma_0)_{\pm1} &= {\textstyle\frac{\text{d}^2 \sigma_{0}}{4}} 
[1-{\textstyle{\frac{1}{\sqrt{3}}}}A_{0}^{0}(11)+\sqrt{{\textstyle\frac{2}{3}}}
A_{0}^{2}(11)],
\end{align}
which represent physically the double differential 
cross-section for $ pp \to pp \omega$ from the initial
spin states $|00 \rangle$, and $|1m \rangle, m=0,\pm1$.
Clearly, measurements of $\sigma_0(\xi), A_{zz}$ and
$A_\Sigma$ are sufficient to determine 
\eqref{d200} to \eqref{d211} individually.

Finally, we may characterize the state of polarization of
the $\omega$  meson  in $pp \to pp \vec{\omega}$ by the density matrix $\rho^s$,
whose elements are given by
\begin{equation}
\rho_{\mu \mu'}^{s} = \frac{2 \pi D}{v}\frac{1}{4}\sum_{s_f}
\sum_{m_f}\langle s\mu;s_f m_f|\boldsymbol{TT}^{\dagger}|s \mu';s_fm_f \rangle.
\end{equation}

Expressing $\rho^s$ in the standard \cite{27} form
\begin{equation}
\rho^s = \frac{1}{2s+1} \sum_{k=0}^{2s} (\tau^k \cdot t^k),
\end{equation}
in terms of $\tau_{\nu}^{k} \equiv S_{\nu}^{k}(s,s)$,the Fano 
statistical tensors $t_{\nu}^{k}$ are given by
\begin{eqnarray}
\label{Fano}
t_{\nu}^{k} &=& \frac{1}{4}\sum_{\alpha,\lambda,\Lambda,
\Lambda'}(-1)^{\lambda -s} [s_f]^2 [s]^3 [\Lambda][\Lambda'] \nonumber \\
& & \times W(s \Lambda s \Lambda';\lambda k)\mathcal{B}_{\nu}^{k}(s_i,s_i),
\end{eqnarray}
at the double differential level. It may be noted that $\rho^s$ is 
unnormalized so that \eqref{Fano} with $k=0$ leads to \eqref{dcs0}. The vector 
and tensor polarizations  of $\omega$ (with $s=1$) are readily obtained by 
setting $k=1,2$ respectively in \eqref{Fano}.

It is worth noting that the Fano statistical
tensors $t_{\nu}^{k}$ may be measured by looking
at the decay  $\omega \to \pi^0 \gamma$ 
\cite{16}, with a branching ratio of  $ 8.92 \% $.  
The angular distribution  of circularly
polarized radiation emitted by polarized  $\omega $   is proportional to 
\begin{equation}
I_p(\theta_{\gamma},\varphi_{\gamma})=\sum_{k=0}^{2}\frac{1}{[k]}
C(11k;p,-p)F_{k}(\theta_{\gamma},\varphi_{\gamma}),
\end{equation}
where  $p=\pm1$ correspond respectively to left and right 
circular polarizations as defined by Rose \cite{26} and
\begin{equation}
F_{k}(\theta_{\gamma},\varphi_{\gamma})=\sum_{q=-k}^{k} (-1)^{q}
t^{k}_{q}Y_{k-q}(\theta_{\gamma},\varphi_{\gamma}),
\end{equation}
where $(\theta_{\gamma},\varphi_{\gamma})$ denote the
polar angles of the direction of $\gamma$ emission in the same frame of
reference in which  $t^{k}_{q}$ are given. If no
observation is made on the polarization of the
radiation, the intensity is proportional to
\begin{equation}
\sum_p I_p(\theta_{\gamma},\varphi_{\gamma})= \sum_{k=0,2}\frac{2}{[k]} 
C(11k;1,-1)F_{k}(\theta_{\gamma},\varphi_{\gamma}),
\end{equation}
from which it is clear  that the tensor polarization can be measured
from the anisotropy of the angular distribution. On the
other hand, the circular polarization asymmetry 
\begin{equation}
\label{cpa}
\Sigma(\theta_{\gamma},\varphi_{\gamma})=
I_{-p}(\theta_{\gamma},\varphi_{\gamma})- 
I_{p}(\theta_{\gamma},\varphi_{\gamma}) =
{\sqrt{2}} F_{1}(\theta_{\gamma},\varphi_{\gamma})
\end{equation}
enables measurement of  vector polarization. 

If the polarization of the $\omega$ meson is measured with 
a nucleon  polarized initially, we may express
\begin{equation}
\label{spfan}
t^{k}_{\nu}= \sum_{\nu'}\mathcal{D}(k,\nu; 1,\nu')P^1_{\nu'}\;;\;k=1,2
\end{equation}
in terms of the  spin transfers
\begin{equation}
\mathcal{D}(k,\nu ; 1,\nu')=
\sum_{\zeta}H\;C(1 \Lambda'' k;\nu'\nu''\nu)
\mathcal{B}^{\Lambda''}_{ \nu''}(s_i,s_i'),
\end{equation}
where $\zeta \equiv (\alpha,
\alpha',\lambda,\lambda',\Lambda,\Lambda',\Lambda'',k')$ and 
\begin{eqnarray}
H &=& -\frac{1}{8}
\sqrt{\frac{3}{2}}(-1)^{\lambda+\lambda'+k'-k} 
[s]^3[s_f]^2[s_i][s_i'][\lambda] \nonumber \\ & \times & 
[\lambda'][\Lambda][\Lambda'][ \Lambda''][k']^2W(s \lambda k'1; \Lambda \lambda')
\nonumber \\ & \times & W(s_i'1s_f \lambda;s_i \lambda')W(1 \Lambda
k \Lambda';k' \Lambda'') \nonumber \\
& \times & W(s \lambda' k \Lambda';k's)
W(s_i'{\textstyle\frac{1}{2}}1{\textstyle\frac{1}{2}};
{\textstyle\frac{1}{2}}s_i),
\end{eqnarray}
if the beam is polarized. If the target is polarized, we may replace 
$P^{1}_{\nu}$ by $Q^{1}_{\nu}$ in \eqref{spfan} and attach a factor
$(-1)^{s_i'-s_i}$ to $H$.

Denoting the six $T^j_{\alpha\beta}$ in Table~\ref{amplitudes} 
serially as $T_1$ to $T_6$, the irreducible tensor amplitudes ${\cal  T}_{\nu}
^{\Lambda}(\alpha,\lambda)$ which describe $pp \to pp \omega $ close to threshold
are explicitly given by
\begin{align}
\label{t1}
{\cal  T}_{\nu}^{1}(101;1) &= \frac{1}{24\pi^{3/2}} T_1 \delta_{\nu0},\\
\label{t2}
{\cal  T}_{\nu}^{1}(100;0) &= \frac{1}{12 \pi} [T_2
+ \frac{3\nu^2-2}{\sqrt{10}}T_3]Y_{1\nu}
(\boldsymbol{\hat{q}}),\\
{\cal  T}_{\nu}^{1}(110;1) &= \frac{1}{12\pi}[T_4
+\frac{3\nu^2-2}{\sqrt{10}}T_5]Y_{1\nu}( \boldsymbol{\hat{p}}_f),\\
\label{t4}
{\cal  T}_{\nu}^{2}(210;1) &= \frac{1}{20 \sqrt{6}\pi}
\nu(4-\nu^2)^{1/2}T_6 Y_{1\nu}(\boldsymbol{\hat{p}}_f),
\end{align}
since $ Y_{l_im_i}(\boldsymbol{\hat{p}}_i)
=([l_i]/\sqrt{4\pi}) \delta_{m_i0} $,
if we choose the beam direction as the z-axis.
All the observables considered in the above discussion are
readily evaluated using \eqref{t1} to \eqref{t4}
in terms of the six partial wave amplitudes   and the angles
characterizing $ {\boldsymbol{q}} $ and ${\boldsymbol{p}}_f$. 
The unpolarized differential cross section measured
in \cite{23} is readily evaluated after integrating 
\eqref{dcs0} with respect to $\text{d}\Omega_{p_f}\text{d}\epsilon$,
where $\epsilon=W-2M$, and we have
\begin{equation}
\label{poly}
\text{d}\sigma_0 = a_0 + a_2 \cos^2 \theta,
\end{equation}
where $a_0$ derives contributions from all the irreducible tensor amplitudes,
while  ${\cal  T}_{\nu}^{1}(100;0)$ alone, which produces the meson in $p-$wave,
contributes to $a_2$. The existing data \cite{23} is in good agreement with 
the form \eqref{poly}, which hence provides clear evidence for the presence of 
the initial spin singlet amplitude ${\cal  T}_{\nu}^{1}(100;0)$ 
given by \eqref{t2} in addition to the initial spin triplet threshold amplitude 
${\cal{T}}^1_\nu(101;1)$ given by \eqref{t1}. If we can assume the contribution 
of ${\cal{T}}^1_\nu(110;1)$ and ${\cal{T}}^2_\nu(210;1)$ to be small or 
negligible, $a_0$ and $a_2$ involve the bilinear combinations
$\,[|T_1|^2+3|T_2+\frac{1}{\sqrt{10}}T_3|^2]$ and 
$[\,|T_3|^2-2 \sqrt{10}\Re(T_2 T_3^*)]$ of the partial wave amplitudes duly 
integrated with respect to $\epsilon$. If one measures not  only the angular 
distribution of $\omega$ but also its energy, the integration with respect to 
$\epsilon$ can be dispensed with.  
 
Integrating the right hand side of \eqref{ak12} with respect to
$\text{d}\Omega_{p_f}$ and equating it to $\text{d}\sigma_0
A^k_{\nu}(k_1 k_2)$  defines the analyzing powers at the $\text{d}^3 q$ level.
It is interesting to note that the Wigner $9j$ symbol in \eqref{wig} ensures that 
the initial spin triplet amplitude \eqref{t1} alone contributes to 
$A^2_0(11)$,  a measurement of which  determines $|T_1|^2$. Knowledge of  $|T_1|^2$
leads to a determination of $|T_2+ \frac{1}{\sqrt{10}}T_3|^2$ using the 
above  expression for $a_0$. Moreover it is interesting to note that 
$\boldsymbol{A}(10) - \boldsymbol{A}(01)$ or  $\boldsymbol{A}(11)$ are proportional
to the interference of the initial spin triplet amplitude ${\cal  T}_{\nu}^{1}(101;1)$
with the initial spin singlet amplitude $ {\cal  T}_{\nu}^{1}(100;0)$.
This leads to a bilinear involving $T_1$ with 
$T_2+\frac{1}{\sqrt{10}}T_3$. Likewise, $t^k_{\nu}$ at $\text{d}^3q$ level are
also obtained on integration of  \eqref{Fano} or \eqref{spfan} with 
respect to $\text{d}\Omega_{p_f}$. There is as yet no data available on any of the spin
observables.  

\begin{acknowledgments}
We thank the referee for helpful suggestions. 
One of us (G.R.) is grateful to Professors B.V. Sreekantan, R. Cowsik and
J.H. Sastry for facilities provided for research at the Indian Institute of
Astrophysics. (M.S.V.) acknowledges support of Department of Science and
Technology, India, (P.N.D.) is thankful to the Alexander von Humboldt
Foundation for the award of a  Fellowship and (J.B.) acknowledges 
encouragement for research given by the Principal Dr. T.G.S. Moorthy and the
Management of K.S. Institute of Technology.
\end{acknowledgments}

\end{document}